\documentclass[12pt]{article}
\def \bea{\begin{eqnarray}}
\def \beq{\begin{equation}}

\def \b{{\cal B}}

\def \ep{\epsilon}

\def \eea{\end{eqnarray}}
\def \eeq{\end{equation}}

\def \Ct{C_\theta}
\def \St{S_\theta}
\def \s{\sqrt{2}}

\def \sx{\sqrt{6}}

\begin{document}

\begin{flushright}
TECHNION-PH-2006-01\\
EFI 06-01 \\
hep-ph/0601129 \\
January 2006 \\
\end{flushright}

\bigskip
\medskip
\begin{center}
\large
{\bf Suppression of Flavor Symmetry Breaking\\
in $B$ Decay Sum Rules
\footnote{Submitted to Physics Letters B.}}

\bigskip
\medskip

\normalsize
{\it Michael Gronau and Yuval Grossman} \\

\medskip
{\it Department of Physics, Technion-Israel Institute of Technology \\
Technion City, Haifa 32000, Israel}

\bigskip
{\it Guy Raz}

\medskip
{\it Department of Physics, Weizmann Institute of Science\\
Rehovot  76100, Israel}

\bigskip
and
\bigskip

{\it Jonathan L. Rosner} \\
\medskip

{\it Enrico Fermi Institute and Department of Physics \\
University of Chicago, Chicago, Illinois 60637} \\

\bigskip
\bigskip

{\bf ABSTRACT}

\end{center}

\begin{quote}

While flavor symmetries are useful for studying hadronic $B$ decays, symmetry
relations for amplitudes and decay rates are usually violated by first order 
symmetry breaking corrections.  We point out two cases in which first order
symmetry breaking is suppressed by a small ratio of amplitudes: (1)  An isospin
sum rule for four $B\to K\pi$ decays, where isospin breaking is shown to be
negligible. (2) An SU(3) sum rule for pairs of $B\to K\pi$ and $B\to K\eta_8$,
generalized to pairs of $B\to K\pi,~B\to K\eta$ and $B\to K\eta'$. 
\end{quote}

Charmless hadronic  $B$ decays provide valuable tests for the pattern of CP 
violation in the Cabibbo-Kobayashi-Maskawa (CKM) framework. Several methods,  
using isospin symmetry and flavor SU(3) symmetry relations for decay
amplitudes, have been developed for extracting CKM phases
\cite{Charles:2004jd,Gronau:2005cz}.  Flavor symmetry relations for amplitudes
are expected to involve corrections which are in most cases linear in symmetry
breaking parameters.  The parameters describing isospin and SU(3) breaking are
respectively of order 
\beq\label{I-SU3br}
\ep_I \equiv \frac{m_d - m_u}{\Lambda_{\rm QCD}} \sim 0.03~~~~~
{\rm and}~~~~~~\delta_{SU(3)} \equiv \frac{m_s-m_d}{\Lambda_{\rm QCD}}
 \sim 0.3~~.
\eeq 
Thus, corrections at corresponding levels have been shown to affect the 
determination of $\alpha$ in $B\to\pi\pi, \rho\pi, \rho\rho$ when applying
methods based on isospin~\cite{Gardner:1998gz,Gronau:2005pq} and flavor 
SU(3)~\cite{Gronau:2004ej,Gronau:2004tm}.

Isospin symmetry has been employed also to derive an approximate sum rule,
relating penguin-dominated $B\to K\pi$ decay rates involving a charged pion
with those containing a neutral pion~\cite{Gronau:1998ep,Matias:2001ch},
\beq\label{SRrateKpi}
\Gamma(B^+ \to K^0 \pi^+) + \Gamma(B^0 \to K^+ \pi^-) \approx 
2[\Gamma(B^+ \to K^+ \pi^0)+\Gamma(B^0 \to K^0 \pi^0)]~~.
\eeq
Terms quadratic in small ratios of amplitudes, which may violate this sum rule,
have been calculated,
implicitly in the isospin symmetry limit, and were found to be
between one and five percent~\cite{Gronau:2003kj,Beneke:2003zv,Bauer:2005kd}.
Corrections of similar order are expected to originate from isospin symmetry
breaking if it is first order, as anticipated on general grounds.  These
corrections must be considered in order to provide evidence for 
physics beyond the Standard Model if a deviation from the sum rule is observed.

The purpose of this Letter is to show that the leading terms in the $B\to K\pi$
sum rule do not involve first order isospin breaking. That is, first order
isospin breaking corrections are suppressed by a small ratio of tree and
penguin amplitudes and may be safely neglected.  We will also consider a sum 
rule combining decay rates for $B\to K\pi$ and $B\to K\eta_8$ (generalized to 
$B\to K\pi,~B\to K\eta$ and $B\to K\eta'$), showing that linear SU(3) breaking
in this sum rule is similarly suppressed by a small ratio of amplitudes.  In
both examples, amplitude quadrangle relations which are exact in the isospin
and SU(3) symmetry limits, respectively, obtain very small symmetry breaking
corrections, comparable  to isospin breaking in subleading tree amplitudes.
A similar situation has been noted ten years ago in a triangle amplitude
relation for $B^+\to \pi^+\pi^0,~B^+\to \pi^+\eta$ and $B^0\to \pi^0\eta$,
where both SU(3) breaking and isospin breaking are suppressed by small ratios
of amplitudes~\cite{Dighe:1995gq}.

We start our discussion with the four $B\to K\pi$ decay
amplitudes. There are three isospin-invariant amplitudes, 
\beq
B_{1/2}\equiv \langle 1/2 | {\cal H}_W^{\Delta I = 0} | 1/2 \rangle,\quad
A_{1/2}\equiv \langle 1/2 | {\cal H}_W^{\Delta I = 1} | 1/2 \rangle, \quad
A_{3/2}\equiv \langle 3/2 | {\cal H}_W^{\Delta I = 1} | 1/2 \rangle.
\eeq
(A fourth one, $C_{3/2}\equiv \langle 3/2 | {\cal H}_W^{\Delta I = 2} |
1/2 \rangle$ vanishes in the isospin limit).
The amplitudes are
\bea\label{iso}
-A(B^0 \to K^+\pi^-) &=& B_{1/2} - A_{1/2} - A_{3/2}~~,\cr
A(B^+ \to K^0\pi^+) &=&  B_{1/2} + A_{1/2} + A_{3/2}~~, \cr
-\s A(B^+ \to K^+\pi^0) &=& B_{1/2} + A_{1/2} - 2A_{3/2}~~,\cr
\s A(B^0 \to K^0\pi^0) &=& B_{1/2} - A_{1/2} + 2A_{3/2}~~.
\eea
This implies a well-known quadrangle amplitude relation which is exact in the
isospin symmetry limit (for compact notation we denote amplitudes from now on
by final states alone)~\cite{Lipkin:1991st}:
\bea \label{SRampKpi}
A(K^0 \pi^+)  -  A(K^+ \pi^-) 
+ \s A(K^+ \pi^0) - \s A(K^0 \pi^0)=0~~.
\eea
We will consider first order isospin breaking in this equation and in
Eq.~(\ref{SRrateKpi}).

It is useful to express the isospin amplitudes in terms of contributions
corresponding to flavor SU(3) topologies for final states involving two members
of the light pseudosdcalar octet~\cite{Gronau:1994rj}, 
\beq\label{flavor}
B_{1/2} = p + {1\over 2}(t + A)~,~~~
A_{1/2} ={1\over 6}(-t +2c + 3A)~,~~~
A_{3/2} = - {1\over 3}(t + c)~~.
\eeq
(In the appendix we perform an equivalent calculation using a tensor language.) 
Each contribution may involve more than one CKM factor.  The terms $p,
t$ and $c$, contain respectively a penguin amplitude $P$, a
``color-favored" tree amplitude $T$, and a ``color-suppressed" tree
amplitude $C$, each appearing in a given linear combination with
either a ``color-favored" ($P_{EW}$) or a ``color-suppressed"
($P^C_{EW}$) electroweak penguin amplitude~\cite{Gronau:1995hn},
\beq
p \equiv P -\frac{1}{3}P^C_{EW}~,~~~
t \equiv T + P^C_{EW}~,~~~
c\equiv C +P_{EW}~~.
\eeq

The above contributions obey a simple hierarchy. The dominant term is the
penguin amplitude 
$P$. Flavor SU(3) and the measured decay rates for $B\to K\pi$ and $
B\to\pi\pi$~\cite{Group(HFAG):2005rb} show that the two tree amplitudes $T$ and $C$ are much 
smaller than $P$. For instance~\cite{Gronau:1994bn,Fleischer:1997um}
\beq
\frac{|T+C|}{|P|} \simeq \frac{\s V_{us}f_K}{V_{ud}f_\pi}
\sqrt{\frac{\b(B^+\to\pi^+\pi^0)}{\b(B^+\to K^0\pi^+)}} = 0.19 \pm 0.02~~.
\eeq
The two electroweak penguin contributions, $P_{EW}$ and $P^C_{EW}$, may be related 
by SU(3) to $T$ and $C$~\cite{Neubert:1998pt}, and are found somewhat smaller than the
latter.
The smallest contribution is the annihilation amplitude $A$, which is expected to be 
suppressed by $1/m_b$ relative to the two tree amplitudes in an approach based on 
Soft Collinear Effective Theory~\cite{Bauer:2004ck},
\beq
\frac{|A|}{|T|} \sim \frac{|A|}{|C|} \sim \frac{\Lambda_{\rm QCD}}{m_b/2} \sim 0.2~~.
\eeq
Thus, we will use
\beq
|A| \ll |t|, |c| \ll |p|~~,
\eeq
where the successive hierarchy suppression factor is about 0.2. 

The isosinglet amplitude $B_{1/2}$ is the only one containing the dominant amplitude $p$,
while the isovector amplitudes $A_{1/2}$ and $A_{3/2}$ are subdominant. 
Eq.~(\ref{SRampKpi}) is obeyed separately for the dominant and subdominant isospin amplitudes.
The vanishing of the three contributions $B_{1/2},~A_{1/2}$ and $A_{3/2}$ in this linear 
combination of hadronic matrix elements holds for {\em any $\Delta I=0$ and 
$\Delta I=1$ operators}. This is the crucial point in understanding 
why first order isospin breaking vanishes in the leading terms in Eq.~(\ref{SRampKpi}).

Consider isospin breaking due to the $d$ and $u$ quark mass difference or due to their charge difference.  
The ``spurion" operator representing all these differences transforms like a sum of 
$\Delta I=0$ and $\Delta I=1$ operators. 
The $\Delta I=0$ operator in the effective Hamiltonian, with the dominant $B\to K\pi$ matrix 
element $B_{1/2}$, becomes 
sum of $\Delta I=0$ and $\Delta I=1$ operators. The linear combination of matrix 
elements appearing in Eq.~(\ref{SRampKpi}) vanishes for these operators  as it would for any 
combination of $\Delta I=0$ and $\Delta I=1$ operators.

To illustrate this general behavior, let us consider first order isospin
breaking effects, which can be visualized diagramatically in terms of quark
mass insertions on quark lines~\cite{Gronau:1995hm}.  We will distinguish
between four types of effects: (i) A spectator effect in $p, t, c$ which we 
denote by parameters $\ep^p_S, \ep^t_S, \ep^c_S$, each representing a sum of
$\Delta I=0$ and $\Delta I=1$ corrections. Thus,  for a spectator 
$d$-quark one has $p_d\equiv p, t_d\equiv t, c_d\equiv c$ while for a spectator
$u$-quark, $p_u \equiv (1+\ep^p_S)p, t_u \equiv (1+\ep^t_S)t, c_u \equiv
(1+\ep^c_S)c$.  
(ii) A $q\bar q$ pair-production (``popping") effect in $p$,
distinguished similarly by a parameter $\ep^p_P$.  (iii) Mixing of $\pi_3\equiv
(d\bar d -u\bar u)/\s$ with an isospin singlet pseudoscalar, $\pi_1 \equiv
(d\bar d + u\bar u)/\s$, given by a $\Delta I = 1$
mixing parameter $\ep_1$, where $\pi^0 =
\pi_3 + \ep_1\pi_1$.  (iv) The isosinglet state $\pi_1$ couples to a new
penguin amplitude $s$ (possibly comparable to $p$) involving a pair-production
(``popping") of a light $q\bar q$ pair which is a singlet under both isospin
and color.  (See discussion below of amplitudes related to the SU(3) singlet
$\eta_1$).  We expect  all five isospin breaking parameters to be roughly equal
in magnitude to $\ep_I$ in Eq.~(\ref{I-SU3br}).

Using these notations, the four physical amplitudes including first order
isospin breaking corrections are:
\bea\label{epKpi}
A(K^0\pi^+) &= &  (1 + \ep^p_S)p + A~~,\nonumber\\
-A(K^+\pi^-) & = & (1 + \ep^p_P)p + t~~,\nonumber\\
-\s A(K^+\pi^0) & = & (1 + \ep^p_S + \ep^p_P -\ep_1)p - \ep_1s +
(1+\ep^t_S-\ep_1)t \nonumber\\
&  & + (1+\ep^c_S-\ep_1)c + A~~, \nonumber\\
\s A(K^0\pi^0) & = &  (1 + \ep_1)p +\ep_1s - (1-\ep_1)c~~.
\eea 
This implies
\beq \label{Ibreak-ampKpi}
A(K^0 \pi^+)  -  A(K^+ \pi^-) + \s A(K^+ \pi^0) - \s A(K^0 \pi^0) = - (\ep^t_S-\ep_1)t - 
\ep^c_Sc~~.
\eeq
We see that the leading first order isospin breaking contributions originating 
in $p$ (and $s$) cancel, as has been argued above. The remaining isospin breaking 
terms are suppressed by $|t|/|p|$ and $|c|/|p|$ relative to these contributions.

Consider now the sum rule (\ref{SRrateKpi}) for $B\to K\pi$ decay rates.
Using Eqs.~(\ref{epKpi}) we note that first order isospin breaking in 
the dominant $|p|^2$ terms (and potential
${\rm Re}(p^*s)$  terms) cancel in this sum rule. 
Omitting a few subleading terms common to both sides and neglecting terms involving 
$A$, the sum rule reads 
(for simplicity of expressions we will assume everywhere that symmetry breaking
parameters are real):
\bea
2|p|^2 = 2|p|^2 & + & 2[|c|^2+ {\rm Re}(c^*t)]  + 2(\ep^p_S + \ep^t_S - 2\ep_1)
{\rm Re}(p^*t)-2\ep_1{\rm Re}(s^*t)\nonumber\\
& + & 2(\ep^p_S + \ep^p_P + \ep^c_S - 2\ep_1){\rm Re}(p^*c)
- 4\ep_1{\rm Re}(s^*c)~~.
\eea
The first terms, $2[|c|^2+ {\rm Re}(c^*t)]$, which violate the sum rule (\ref{SRrateKpi}) in the 
isospin symmetry limit, have been estimated to correspond to a ratio $[|c|^2+ {\rm Re}(c^*t)]/|p|^2$ 
in the range $1-5\%$~\cite{Gronau:2003kj,Beneke:2003zv,Bauer:2005kd}.
The remaining subleading isospin breaking terms, of forms 
$2\ep_I{\rm Re}(p^*t),~2\ep_I{\rm Re}(p^*c),~2\ep_I{\rm Re}(s^*t)$ and $2\ep_I{\rm Re}(s^*c)$,
are expected to be smaller by about a factor $\ep_I|p|/|c| \sim 0.15$ 
relative to these terms. The neglected terms involving $A$ are even smaller. Therefore, 
isospin breaking corrections in the sum rule are negligible in the Standard Model.

We now turn to an SU(3) quadrangle relation for  amplitudes involving pairs of $B\to K\pi$ 
and $B\to K\eta_8$ decays:
\beq\label{Keta8}
A(K^+\pi^-) + A( K^0\pi^+) + \sx A(K^0\eta_8) - \sx A(K^+\eta_8) = 0~~.
\eeq
This relation reads, in terms of graphical contributions representing SU(3) 
amplitudes~\cite{Gronau:1994rj,Gronau:1995hn},
\beq
[-p-t] + [p+A] + [p-c] -[p-c-t+A] = 0 ~~.
\eeq

First order SU(3) breaking caused by the $s$ and $d$ (or $u$) quark mass-difference 
occurs in the dominant amplitude $p$ and in the doubly suppressed amplitude $A$,
while there is no SU(3) breaking in $t$ and $c$. Symmetry breaking in $p$ and $A$  
is due to $s\bar s$ 
popping in $B\to K \eta_8$ versus $u\bar u$ or $d\bar d$ popping in $B\to K\pi$. 
We denote penguin and annihilation amplitudes by the popping $q\bar q$ pair, 
using $\delta^p$ and $\delta^A$ to describe SU(3) breaking in $p$ and $A$.
Thus, one has $p_u=p_d\equiv p,~p_s = (1+ \delta^p)p, A_u=A_d\equiv A,~A_s=
(1+\delta^A)A$, where one expects 
$|\delta^p| \sim |\delta^A| \sim 
\delta_{SU(3)}\sim 0.3$. We find that terms multiplying $p$ involving the
parameter $\delta^p$ cancel in Eq.~(\ref{Keta8}),
\beq
[-p-t] + [p+A] + [(1+2\delta^p)p-c] -[(1+2\delta^p)p-c-t+(1+2\delta^A)A] = 2\delta^AA ~~.
\eeq
Higher order SU(3) breaking corrections in $p$ caused by the $s$ and $d$ (or $u$) quark 
mass-difference are given by an arbitrary number of $s\bar s$ insertions. Since $s\bar s$ 
and the operator contributing to $p$ are both $I=0$ operators, SU(3) breaking in $p$ from this source alone {\em without isospin breaking} cancels in (\ref{Keta8}) to all orders.
 
The remaining first order SU(3) breaking correction proportional to $\delta^A$ is suppressed by 
$|A|/|p|\sim (0.2)^2$ relative to each of the four amplitudes which are dominated by $p$. At this 
low level the corrections from first order isospin breaking effects could be of the same order.
Including isospin breaking spectator corrections in $p, t$ and $c$ as discussed above,  
Eq.~(\ref{Keta8}) now reads:
\bea\label{Keta8ampBrake}
A(K^+\pi^-) +  A( K^0\pi^+) & + & \sx A(K^0\eta_8) - \sx A(K^+\eta_8) \nonumber\\
& = & 2\delta^AA -2\delta^p\ep^p_Sp + \ep^t_St +\ep^c_Sc~~.
\eea
The right-hand side consists of four terms representing first order SU(3) breaking in $A$, first 
order SU(3) breaking and first order isospin breaking in $p$, and first order isospin breaking in $t$ and
$c$. All four terms are expected to be of comparable magnitudes and are comparable to the isospin 
breaking corrections appearing in the $K\pi$ quadrangle relation Eq.~(\ref{Ibreak-ampKpi}).

The decay rates of the four processes in Eq.~(\ref{Keta8}) obey an approximate sum rule,
relating processes involving  a neutral kaon with those containing a charged kaon:
\beq\label{Keta8rate}
\Gamma(K^0\pi^+) + 6\Gamma(K^0\eta_8)  \approx 
\Gamma(K^+\pi^-) + 6\Gamma(K^+\eta_8)~~.
\eeq
In terms of SU(3) invariant amplitudes and including leading SU(3) breaking corrections this 
reads, after omitting subleading terms which are common to both sides:
\beq\label{Keta8rateBrake}
 2|p|^2 = 2[|p|^2 + |t|^2 + {\rm Re}(c^*t) - 2\delta^p{\rm Re}(p^*t)]~~.
 \eeq
Indeed, first order SU(3) breaking corrections in the dominant $|p|^2$ term cancel in (\ref{Keta8rate}). 
The remaining SU(3) breaking in the subleading term ${\rm Re}(p^*t)$ is 
comparable to the two quadratic subleading terms, $|t|^2$ and ${\rm Re}(c^*t)$, which 
violate the sum rule in the SU(3) symmetry limit. 

So far we have considered decays to states involving the unphysical SU(3) octet state 
$\eta_8$. SU(3) breaking mixes the $\eta_8$ with the SU(3) singlet $\eta_1$ in the physical 
$\eta$ and $\eta'$. In order to discuss relations for the physical states analogous to  
Eqs.~(\ref{Keta8}) and (\ref{Keta8rate}) and to study their SU(3) breaking [as we did in Eqs.~(\ref{Keta8ampBrake}) and (\ref{Keta8rateBrake}) for the unphysical case], we use the
single mixing angle parametrization,
\beq
\eta = \cos\theta\eta_8 + \sin\theta\eta_1~~,~~~~~~
\eta' = \cos\theta\eta_1 - \sin\theta\eta_8~~.
\eeq
The measured value~\cite{Escribano:2005qq}, $\sin\theta \simeq -1/3$, has a magnitude 
characteristic 
of SU(3) breaking [see Eq.~(\ref{I-SU3br})]. 
In the following discussion of processes involving $\eta$ and $\eta'$ we will keep the explicit dependence on $\theta$. 

$B$ decays to final states with a singlet $\eta_1$ and an octet pseudoscalar meson are  
described by three SU(3) amplitudes, corresponding to operators in the effective Hamiltonian 
transforming under SU(3) as $3, \overline{6}$ and $15$~\cite{Dighe:1995bm}. 
The graphical representations of these three amplitudes are in terms of a ``singlet" 
penguin amplitude, $s\equiv S - P^C_{EW}/3$,  involving a popping $q\bar q$ 
pair which is a singlet under both flavor and color, and annihilation and exchange 
amplitudes to which an $\eta_1$ is connected by two gluons attached to quark lines
in all possible ways. The annihilation and exchange amplitudes, in which the two gluons are 
attached to the spectator quark are topologically equivalent to tree amplitudes, 
$t_s$ and $c_s$. For instance, in $t_s$ a spectator $u$ and a $\bar u$ from $\bar b\to \bar u$ 
are connected to $\eta_1$ by two gluons. The singlet annihilation and exchange amplitudes, 
which have been often neglected~\cite{Dighe:1995bm,Chiang:2003rb}, have recently
been shown to be of leading order in $1/m_b$~\cite{Williamson:2006hb}
and will be included here. Of these 
two amplitudes only  $t_s$ contributes to strangeness changing $B^0$ and $B^+$ decays.
We expect $s$ to be dominant and $t_s$ to be subdominant in $B\to K \eta_1$,
similar to $p$ and $t$ in $B\to K\pi$.

Neglecting the $1/m_b$-suppressed contribution $A$, we write expressions 
for amplitudes  in the SU(3) symmetry limit while keeping $\theta$ as the 
only SU(3) breaking parameter (we denote 
$S_\theta \equiv \sin\theta, C_\theta \equiv \cos\theta$)~\cite{Dighe:1995bm}:
\bea\label{ampetaeta'}
\sx A(K^+\eta) & = & (\Ct+2\s\St)p + 3\s\St s -(\Ct -\s\St) c -\Ct t + \s\St(t+t_s)~~,\nonumber\\
\sx A(K^0\eta) & = & (\Ct+2\s\St)p + 3\s\St s -(\Ct -\s\St) c~~,\nonumber\\
\sx A(K^+\eta') & = & (2\s\Ct-\St)p + 3\s\Ct s + (\s\Ct+\St)c +\s\Ct(t+t_s) + \St t~~,\nonumber\\
\sx A(K^0\eta') & = & (2\s\Ct-\St)p + 3\s\Ct s + (\s\Ct+\St)c~~.
\eea
This implies an amplitude relation,
\beq
\sx\left (\Ct[A(K^+\eta) - A(K^0\eta)] -\St[A(K^+\eta') - A(K^0\eta')] \right )= 
A(K^0\pi^+) + A(K^+\pi^-)~~,
\eeq
which becomes Eq.~(\ref{Keta8}) in the SU(3) symmetry limit, $\theta=0$. 

Taking the physical $\eta$ to replace $\eta_8$ in~(\ref{Keta8rate}) the
approximation
\beq\label{Ketarate}
\Gamma(K^0\pi^+) + 6\Gamma(K^0\eta)  \approx 
\Gamma(K^+\pi^-) + 6\Gamma(K^+\eta)~~.
\eeq
now reads, with the same omissions on both sides as in (\ref{Keta8rateBrake})
and to first order in $SU(3)$ breaking,
\begin{equation}
  \label{KetarateBrake}
  2|p|^2 = 2[|p|^2 + |t|^2 + {\rm Re}(c^*t) -
  (2\delta^p+\sqrt{2}S_\theta){\rm Re}(p^*t) + \s\St{\rm Re}(p^*t_s)
  - 3\,\sqrt{2} S_{\theta}\,{\rm Re}(s^*t)]~~.
\end{equation}
Flavor SU(3) studies of $B$ decays into two pseudoscalars have shown that 
$|s| < |p|$~\cite{Chiang:2003rb}, and we have $|S_\theta| \sim |\delta^p|$. 
Therefore this approximation is qualitatively similar to~(\ref{Keta8rateBrake}). 

Finally, we turn to a sum rule obtained from Eqs.~(\ref{ampetaeta'}) 
for decay rates involving $K\pi,~K\eta$ and $K\eta'$,
\beq\label{Ketaeta'rate}
\frac{\Gamma(K^0\eta)}{1- \s\tan\theta} + \frac{\Gamma(K^0\eta')}{1+
\s\cot\theta} +\frac{\Gamma(K^0\pi^+)}{6} \approx
\frac{\Gamma(K^+\eta)}{1- \s\tan\theta} + \frac{\Gamma(K^+\eta')}{1+
\s\cot\theta} +\frac{\Gamma(K^+\pi^-)}{6}~~.
\eeq
While the leading terms on both sides include dominant terms $|p|^2, |s|^2$ and
${\rm Re}(p^*s)$, we neglect on the right-hand side much smaller terms, $|t|^2$
and ${\rm Re}(c^*t)$, 
as we did in Eqs.~(\ref{Keta8rate}) and (\ref{Ketarate}).  In fact, in the
SU(3) symmetry limit (including $\theta =0$) Eq.~(\ref{Ketaeta'rate}) becomes identical
to Eq.~(\ref{Keta8rate}). SU(3) breaking corrections in the above three
dominant terms cancel. 
First order SU(3) breaking corrections occuring in subleading terms in the sum rule 
(\ref{Ketaeta'rate}) are 
$(2/3)\delta^p{\rm Re}(p^*t) + \s\St{\rm Re}(p^*t_s) + \s\St{\rm Re}(s^*t_s) $. 
These terms are formally similar in magnitude to the
SU(3) breaking corrections in Eq.~(\ref{KetarateBrake}). Since $|s|<|p|$~\cite{Chiang:2003rb},
they would be smaller if also $|t_s|<|t|$, as has been often assumed. 

\begin{table}
\caption{Branching ratios \cite{Group(HFAG):2005rb,Abe:2005iy} entering into
tests of decay rate sum rules.
\label{tab:brs}}
\begin{center}
\begin{tabular}{c c} \hline \hline
Decay   & ${\cal B}$ \\
process & (units of $10^{-6}$) \\ \hline
$B^+ \to K^0 \pi^+$ & $24.1 \pm 1.3$ \\
$B^+ \to K^+ \pi^0$ & $12.1 \pm 0.8$ \\
$B^0 \to K^+ \pi^-$ & $18.9 \pm 0.7$ \\
$B^0 \to K^0 \pi^0$ & $11.5 \pm 1.0$ \\
$B^+ \to K^+ \eta$  &  $2.5 \pm 0.3$ \\
$B^0 \to K^0 \eta$  & $0.9 \pm 0.6 \pm 0.1^{~a}$\\
$B^+ \to K^+ \eta'$ & $69.4 \pm 2.7$ \\ 
$B^0 \to K^0 \eta'$ & $63.2 \pm 3.3$ \\ \hline \hline
\end{tabular}
\end{center}
\leftline{\hskip 1.8in $^a$ Reference \cite{Abe:2005iy}}
\end{table}

We conclude with a brief discussion of the present status of the sum rules
(\ref{SRrateKpi}), (\ref{Ketarate}) and (\ref{Ketaeta'rate}).  Similar sum rules are obeyed
by $B^+$ and $B^0$ branching ratios if one corrects for the $B^+/B^0$
lifetime ratio \cite{Group(HFAG):2005rb} $\tau_+/\tau_0 = 1.076 \pm 0.008$, for
instance by multiplying all $B^0$ branching ratios by this ratio.  We shall
adopt this procedure.

Using the latest averages for branching ratios \cite{Group(HFAG):2005rb} quoted
in Table \ref{tab:brs}, one finds that, in units of $10^{-6}$, the sum
rule (\ref{SRrateKpi}) may be written
\beq
44.4 \pm 1.5 = 48.9 \pm 2.7~~~,
\eeq
which is violated at the level of $1.5 \sigma$.  More precise data are needed in 
order to test the Standard Model expectation 
\cite{Gronau:2003kj,Beneke:2003zv,Bauer:2005kd} that the sum rule should hold 
to an accuracy of between one and five percent.
The isospin breaking effects discussed here were shown to be considerably smaller.  

The sum rule (\ref{Ketarate}) reads
\beq
29.9 \pm 4.1 = 35.3 \pm 2.0~~~,
\eeq
where the error on the left-hand side is dominated by the uncertainty
in ${\cal B}(B^0 \to K^0 \eta)$. 
The sum rule is satisfied and there is no evidence for
large SU(3) breaking effects.  Indeed, we did anticipate in
Eq.~(\ref{KetarateBrake}) a suppression of these effects by ${\cal
O}(|t/p|)$.

For $\theta = {\rm sin}^{-1}(-1/3) \simeq -19.47^\circ$, consistent with
measurements \cite{Escribano:2005qq}, the sum rule (\ref{Ketaeta'rate}) reads
\beq
\frac{2}{3} \Gamma(K^0 \eta) - \frac{1}{3} \Gamma(K^0 \eta') + \frac{1}{6}
\Gamma(K^0 \pi^+)=
\frac{2}{3} \Gamma(K^+ \eta) - \frac{1}{3} \Gamma(K^+ \eta') + \frac{1}{6}
\Gamma(K^+ \pi^-) ~~,
\eeq
\beq
{\rm or}~~~-18.0 \pm 1.3 = -18.1 \pm 0.9~~~.
\eeq
Here the experimental errors on the $K \eta'$ branching ratios are the dominant
source of uncertainties.
As we have argued, SU(3) breaking corrections in this sum rule are suppressed,
occurring only  in subleading terms. 
Assuming $|t_s|<|t|$, one would expect these 
corrections to be smaller than those affecting the sum rule (\ref{Ketarate}).
 
\medskip
We thank Yossi Nir and Jure Zupan for helpful discussions and wish
to congratulate Jure for the birth of Tilen.~J. L. R. wishes to thank
the Technion -- Israel Institute of Technology for gracious
hospitality during part of this work.  This work was supported in part
by the United States Department of Energy through Grant No.\ DE FG02
90ER40560, by the Israel Science Foundation under Grants No. 1052/04
and 378/05, and by the German--Israeli Foundation under Grant
No. I-781-55.14/2003.

\appendix

\section{Symmetry breaking with isospin invariants}
\label{sec:isosp-break-tensor}

In this work we have used the language of amplitude topologies to discuss
isospin and $SU(3)$ breaking effects. 
The same discussion can be conducted also in
a language of isospin and $SU(3)$ invariant amplitudes. This approach isolates
isospin and SU(3) breaking terms transforming as an isospin triplet and SU(3)
octet, respectively.  This isolation may also be achieved in the graphical
approach. In the case of isospin breaking the spectator effect is then given by
$a_u\equiv (1+\ep^a_S)a$, $a_d\equiv (1-\ep_S^a)a$ for $a=p,~t,~c$ and similar
expressions apply to the ``popping'' effect in $p$.   

In this Appendix we derive the results related to the isospin sum rule again,
using symmetry invariants language.  Using tensor language
we consider the isospin invariant tensors 
\beq
K_i=(K^+\; K^0),\qquad \Pi^i_j
= \left( \begin{array}{cc} -\pi^0/\sqrt{2} & \pi^+ \\ -\pi^- &
    \pi^0/\sqrt{2} \end{array} \right),\qquad 
B^i = \left(\begin{array}{c}
    B^+ \\ B^0 \end{array} \right),
\eeq
 together with the Hamiltonian
singlet operator $H_0$ and the Hamiltonian triplet operator
${H_1}^i_j$. To these ingredients we now add the isospin breaking
triplet which can most generally be parametrized by 
\beq
M^i_j=
\left(\begin{array}{cc} -\epsilon/\sqrt{2} & 0 \\ 0 &
    \epsilon/\sqrt{2} \end{array}\right),
\eeq
 with $\epsilon \sim
\epsilon_I$ of (\ref{I-SU3br}). By considering every possible
contraction of the invariant tensors indices above, one can build the
effective Hamiltonian for all $B \to K \pi$ decays \cite{Savage:1989ub}.

Considering first the contraction of the singlet Hamiltonian operator
$H_0$ with $M^i_j$. The combination is obviously an isospin triplet
$H_0 M^i_j$. Upon contracting with the meson tensors, 
which are combined to create an $I=1/2$ and an $I=3/2$ representation, 
two reduced matrix elements result. 
(More technical details can be found in~\cite{Raz:2005hu}.)
We denote these reduced matrix elements by $A_{1/2}^{\epsilon,0}$ and
$A_{3/2}^{\epsilon,0}$, where the subscripts indicate the $K \pi$
representation and the superscript reminds us that these terms are a
result of combining the isospin breaking triplet $M^i_j$ with the
singlet Hamiltonian operator $H_0$. Those reduced matrix elements
transform 
as $A_{1/2}$ and $A_{3/2}$ respectively. In the same
way we have $B_{1/2}^{\epsilon,1}$, which comes from the
singlet combination
${H_1}^i_jM^j_i$. In principle, there could also be
$A_{1/2}^{\epsilon,1}$ and $A_{3/2}^{\epsilon,1}$, which come from
the triplet combination 
${H_1}^i_kM^k_j$. However, since both $H_1$ and
$M$ have only an $I_z=0$ component the $I=0$ combination of them is
zero. There is one additional reduced matrix element which transforms as
a $\Delta I=2$ term
under isospin. We denote it by $C_{3/2}^{\epsilon,1}$,
coming from the combination ${H_1}^i_jM^k_l$.

Using these matrix elements, we write the four $B\to K \pi$ amplitudes as
\begin{eqnarray}
  A(B^+ \to K^0 \pi^+) & = &
  B_{1/2}+A_{1/2}+A_{3/2} + B_{1/2}^{\epsilon,1} +
  A_{1/2}^{\epsilon,0}  + A_{3/2}^{\epsilon,0} 
  + C_{3/2}^{\epsilon,1} \;, \nonumber \\
  -A(B^0 \to K^+ \pi^-) & = &
  B_{1/2}-A_{1/2}-A_{3/2} + B_{1/2}^{\epsilon,1} -
  A_{1/2}^{\epsilon,0}  - A_{3/2}^{\epsilon,0} 
  + C_{3/2}^{\epsilon,1} \;, \nonumber \\
  -\sqrt{2}\,A(B^+ \to K^+ \pi^0) & = &
  B_{1/2}+A_{1/2}-2\, A_{3/2}  + B_{1/2}^{\epsilon,1} +
  A_{1/2}^{\epsilon,0}  - 2\, A_{3/2}^{\epsilon,0} 
  - 2\,C_{3/2}^{\epsilon,1} \;, \nonumber \\
  \sqrt{2}\,A(B^0 \to K^0 \pi^0) & = &
  B_{1/2}-A_{1/2}+2\, A_{3/2} + B_{1/2}^{\epsilon,1} -
  A_{1/2}^{\epsilon,0}  + 2\, A_{3/2}^{\epsilon,0} 
  - 2\,C_{3/2}^{\epsilon,1} \;.
\end{eqnarray}
The amplitude sum rule  now reads
\begin{equation}
  A(K^0\pi^+)-A(K^+\pi^-)+\sqrt{2}\,A(K^+\pi^0)-\sqrt{2} \, A(K^0
  \pi^0) = 6\, C_{3/2}^{\epsilon,1}\;.
\end{equation}
While from the group theoretical point of view, the (now seven) reduced matrix
elements are all independent,
the transformation properties of the symmetry breaking terms lead us to
expect the approximate relations
\begin{equation}
\label{eq:approx}
B_{1/2}^{\epsilon,1} \sim \epsilon\,A_{1/2}\;,\qquad
A_{1/2}^{\epsilon,0},\,A_{3/2}^{\epsilon,0} \sim \epsilon\,B_{1/2}\;,
\qquad C_{3/2}^{\epsilon,1} \sim \epsilon\,A_{3/2}\;.  
\end{equation}
Since the dominant penguin term is in $B_{1/2}$ we see that the
isospin breaking in the amplitude sum rule is suppressed.

We next write the relation for the rates, dropping terms of
$\mathcal{O}(\epsilon^2)$. We get
\begin{eqnarray*}
  && \Gamma(K^0\pi^+)+\Gamma(K^+\pi^-)-2\,\Gamma(K^+\pi^0)-2 \,
 \Gamma(K^0 \pi^0)  = 12\, {\rm Re} (A_{3/2}^*
  A_{1/2}) - 6\, \left|A_{3/2}\right|^2  \\
  && \qquad +  12\, {\rm Re} (A_{1/2}^*
  A_{3/2}^{\epsilon,0}) + 12\, {\rm Re} (A_{3/2}^* 
  A_{1/2}^{\epsilon,0}) -12\, {\rm Re} (A_{3/2}^*  A_{3/2}^{\epsilon,0}) + 12\,
  {\rm Re} (B_{1/2}^* C_{3/2}^{\epsilon,1})\;.
\end{eqnarray*}
A linear breaking term in the dominant amplitudes would have been of
order $\sim \epsilon\, |B_{1/2}|^2$. Such a term can only come from 
either ${\rm Re}(B_{1/2}^* A_{1/2}^{\epsilon,0})$ or ${\rm
  Re}(B_{1/2}^* A_{3/2}^{\epsilon,0})$. We see that no such terms exist.

We have also performed a similar analysis for the SU(3) sum rule of
Eq. (\ref{Keta8}). Technical details for such a calculation can be
found in \cite{Raz:2005hu}. The results serve as a check to those
derived using the graphical method as described above.


\begin{thebibliography}{99}

%1
%\cite{Charles:2004jd}
\bibitem{Charles:2004jd}
 J.~Charles {\it et al.}  [CKMfitter Group],
  %``CP violation and the CKM matrix: Assessing the impact of the asymmetric  B
  %factories,''
 Eur.\ Phys.\ J.\ C {\bf 41}, 1 (2005);
 %[arXiv:hep-ph/0406184];
 %%CITATION = HEP-PH 0406184;
 updated in {\tt www.slac.stanford.edu/xorg/ckmfitter.}
 
  %2
%\cite{Gronau:2005cz}
\bibitem{Gronau:2005cz}
 M.~Gronau,
 %``Weak phases and CP violation,''
 arXiv:hep-ph/0510153; talk given at the Tenth International Conference on $B$
 Physics at Hadron Machines, Assisi, Perugia, Italy, June 20--24, 2005.
 %%CITATION = HEP-PH 0510153;%%

  %3
  %\cite{Gardner:1998gz}
\bibitem{Gardner:1998gz}
  S.~Gardner,
  %``How isospin violation mocks *new* physics:  $\pi^0 - \eta, \eta^\prime$
  %mixing in $B \to \pi \pi$ decays,''
  Phys.\ Rev.\ D {\bf 59}, 077502 (1999)
  [arXiv:hep-ph/9806423];
  %%CITATION = HEP-PH 9806423;%%
  %\cite{Gardner:2005pq}
%\bibitem{Gardner:2005pq}
  %S.~Gardner,
  %``Towards a precision determination of alpha in B $\to$ pi pi decays,''
  Phys.\ Rev.\ D {\bf 72}, 034015 (2005)
  [arXiv:hep-ph/0505071].
  %%CITATION = HEP-PH 0505071;%%
  
  %4
  %\cite{Gronau:2005pq}
\bibitem{Gronau:2005pq}
  M.~Gronau and J.~Zupan,
  %``Isospin-breaking effects on alpha extracted in B $\to$ pi pi, rho rho, rho
  %pi,''
  Phys.\ Rev.\ D {\bf 71}, 074017 (2005)
  [arXiv:hep-ph/0502139].
  %%CITATION = HEP-PH 0502139;%%
  
  %5
  %\cite{Gronau:2004ej}
\bibitem{Gronau:2004ej}
  M.~Gronau and J.~L.~Rosner,
  %``Implications of CP asymmetries in B $\to$ pi+ pi-,''
  Phys.\ Lett.\ B {\bf 595}, 339 (2004)
  [arXiv:hep-ph/0405173].
  %%CITATION = HEP-PH 0405173;%%
  
  %6
  %\cite{Gronau:2004tm}
\bibitem{Gronau:2004tm}
  M.~Gronau and J.~Zupan,
  %``On measuring alpha in B(t) $\to$ rho+- pi-+,''
  Phys.\ Rev.\ D {\bf 70}, 074031 (2004)
  [arXiv:hep-ph/0407002].
  %%CITATION = HEP-PH 0407002;%%
  
  %7
  %\cite{Gronau:1998ep}
\bibitem{Gronau:1998ep}
  M.~Gronau and J.~L.~Rosner,
  %``Combining CP asymmetries in B $\to$ K pi decays,''
  Phys.\ Rev.\ D {\bf 59}, 113002 (1999)
  [arXiv:hep-ph/9809384];
  %%CITATION = HEP-PH 9809384;%%
%\cite{Lipkin:1998ie}
%\bibitem{Lipkin:1998ie}
  H.~J.~Lipkin,
  %``A useful approximate isospin equality for charmless strange B decays,''
  Phys.\ Lett.\ B {\bf 445}, 403 (1999)
  [arXiv:hep-ph/9810351].
  %%CITATION = HEP-PH 9810351;%%
  
  %8
  %\cite{Matias:2001ch}
\bibitem{Matias:2001ch}
  J.~Matias,
  %``Model independent sum rules for B $\to$ pi K decays,''
  Phys.\ Lett.\ B {\bf 520}, 131 (2001)
  [arXiv:hep-ph/0105103].
  %%CITATION = HEP-PH 0105103;%%
  In contrast to our language, isospin symmetry and isospin breaking are
  referred to in this paper as weak interaction operators transforming as
  $\Delta I = 0$ and $\Delta I\ne 0$, respectively.  

  %9
%\cite{Gronau:2003kj}
\bibitem{Gronau:2003kj}
  M.~Gronau and J.~L.~Rosner,
  %``Rates and asymmetries in B $\to$ K pi decays,''
  Phys.\ Lett.\ B {\bf 572}, 43 (2003)
  [arXiv:hep-ph/0307095].
  %%CITATION = HEP-PH 0307095;%%
  
  %10
 % \cite{Beneke:2003zv}
\bibitem{Beneke:2003zv}
  M.~Beneke and M.~Neubert,
  %``QCD factorization for B $\to$ P P and B $\to$ P V decays,''
  Nucl.\ Phys.\ B {\bf 675}, 333 (2003)
  [arXiv:hep-ph/0308039].
  %%CITATION = HEP-PH 0308039;%%
  
  %11
  %\cite{Bauer:2005kd}
\bibitem{Bauer:2005kd}
  C.~W.~Bauer, I.~Z.~Rothstein and I.~W.~Stewart,
  %``SCET analysis of B $\to$ K pi, B $\to$ K anti-K, and B $\to$ pi pi
  %decays,''
  arXiv:hep-ph/0510241.
  %%CITATION = HEP-PH 0510241;%%
  
  %12
  %\cite{Dighe:1995gq}
\bibitem{Dighe:1995gq}
  A.~S.~Dighe, M.~Gronau and J.~L.~Rosner,
  %``Amplitude relations for $B$ decays involving $\eta$ and $\eta'$,''
  Phys.\ Lett.\ B {\bf 367}, 357 (1996)
  [Erratum-ibid.\ B {\bf 377}, 325 (1996)]
  [arXiv:hep-ph/9509428].
  %%CITATION = HEP-PH 9509428;%%
  
   %13
%\cite{Lipkin:1991st}
\bibitem{Lipkin:1991st}
  H.~J.~Lipkin, Y.~Nir, H.~R.~Quinn and A.~Snyder,
  %``Penguin trapping with isospin analysis and CP asymmetries in B decays,''
  Phys.\ Rev.\ D {\bf 44}, 1454 (1991);
  %%CITATION = PHRVA,D44,1454;%%
%\cite{Gronau:1991dq}
%\bibitem{Gronau:1991dq}
  M.~Gronau,
  %``Elimination of penguin contributions to CP asymmetries in B decays through
  %isospin analysis,''
  Phys.\ Lett.\ B {\bf 265}, 389 (1991).
  %%CITATION = PHLTA,B265,389;%%

%14
%\cite{Gronau:1994rj}
\bibitem{Gronau:1994rj}
  M.~Gronau, O.~F.~Hernandez, D.~London and J.~L.~Rosner,
  %``Decays of B mesons to two light pseudoscalars,''
  Phys.\ Rev.\ D {\bf 50}, 4529 (1994)
  [arXiv:hep-ph/9404283].
  %%CITATION = HEP-PH 9404283;%%

%15
  %\cite{Gronau:1995hn}
\bibitem{Gronau:1995hn}
 M.~Gronau, O.~F.~Hernandez, D.~London and J.~L.~Rosner,
  %``Electroweak penguins and two-body B decays,''
 Phys.\ Rev.\ D  {\bf 52}, 6374 (1995)
  [arXiv:hep-ph/9504327].
  %%CITATION = HEP-PH 9504327;%%
 
   %16
  %\cite{Group(HFAG):2005rb}
\bibitem{Group(HFAG):2005rb}
K. Anikeev {\it et al.}, Heavy Flavor Averaging Group,
  ``Averages of b-hadron properties as of winter 2005,
  arXiv:hep-ex/0505100.
Updated results and references are tabulated periodically by this group:
{\tt http://www.slac.stanford.edu/xorg/hfag/rare.}
  %%CITATION = HEP-EX 0505100;%%
  
  %17
  %\cite{Gronau:1994bn}
\bibitem{Gronau:1994bn}
  M.~Gronau, J.~L.~Rosner and D.~London,
  %``Weak coupling phase from decays of charged B mesons to pi K and pi pi,''
  Phys.\ Rev.\ Lett.\  {\bf 73}, 21 (1994)
  [arXiv:hep-ph/9404282].
  %%CITATION = HEP-PH 9404282;%%
  
   %18
  %\cite{Fleischer:1997um}
\bibitem{Fleischer:1997um} See also
  R.~Fleischer and T.~Mannel,
  %``Constraining the CKM angle gamma and penguin contributions through
  %combined B $\to$ pi K branching ratios,''
  Phys.\ Rev.\ D {\bf 57}, 2752 (1998)
  [arXiv:hep-ph/9704423];
  %%CITATION = HEP-PH 9704423;%%
  %\cite{Gronau:1997an}
%\bibitem{Gronau:1997an}
  M.~Gronau and J.~L.~Rosner,
  %``Weak phase gamma from ratio of B $\to$ K pi rates,''
  Phys.\ Rev.\ D {\bf 57}, 6843 (1998)
  [arXiv:hep-ph/9711246];
  %%CITATION = HEP-PH 9711246;%%
  %\cite{Neubert:1998re}
%\bibitem{Neubert:1998re}
  M.~Neubert,
  %``Model-independent analysis of B $\to$ pi K decays and bounds on the weak
  %phase gamma,''
  JHEP {\bf 9902}, 014 (1999)
  [arXiv:hep-ph/9812396].
  %%CITATION = HEP-PH 9812396;%
  
  %19
  %\cite{Neubert:1998pt}
\bibitem{Neubert:1998pt} 
  M.~Neubert and J.~L.~Rosner,
  %``New bound on gamma from B+- $\to$ pi K decays,''
  Phys.\ Lett.\ B {\bf 441}, 403 (1998)
  [arXiv:hep-ph/9808493];
  %%CITATION = HEP-PH 9808493;%%
%\cite{Gronau:1998fn}
%\bibitem{Gronau:1998fn}
  M.~Gronau, D.~Pirjol and T.~M.~Yan,
  %``Model-independent electroweak penguins in B decays to two  pseudoscalars,''
  Phys.\ Rev.\ D {\bf 60}, 034021 (1999)
  [Erratum-ibid.\ D {\bf 69}, 119901 (2004)]
  [arXiv:hep-ph/9810482].
  %%CITATION = HEP-PH 9810482;%%
  
  %20
  %\cite{Bauer:2004ck}
\bibitem{Bauer:2004ck}
  C.~W.~Bauer and D.~Pirjol,
  %``Graphical amplitudes from SCET,''
  Phys.\ Lett.\ B {\bf 604}, 183 (2004);
  [arXiv:hep-ph/0408161];
  %%CITATION = HEP-PH 0408161;%%
  %\cite{Bauer:2004tj}
%\bibitem{Bauer:2004tj}
  C.~W.~Bauer, D.~Pirjol, I.~Z.~Rothstein and I.~W.~Stewart,
  %``B $\to$ M(1) M(2): Factorization, charming penguins, strong phases, and
  %polarization,''
  Phys.\ Rev.\ D {\bf 70}, 054015 (2004).
  [arXiv:hep-ph/0401188].
  %%CITATION = HEP-PH 0401188;%%
  
  %21
  %\cite{Gronau:1995hm}
\bibitem{Gronau:1995hm}
  M.~Gronau, O.~F.~Hernandez, D.~London and J.~L.~Rosner,
  %``Broken SU(3) symmetry in two-body B decays,''
  Phys.\ Rev.\ D {\bf 52}, 6356 (1995)
  [arXiv:hep-ph/9504326].
  %%CITATION = HEP-PH 9504326;%%
  
  %22
  %\cite{Escribano:2005qq}
\bibitem{Escribano:2005qq} 
For a recent update on $\eta-\eta'$ mixing, see
  R.~Escribano and J.~M.~Frere,
  %``Study of the eta eta' system in the two mixing angle scheme,''
  JHEP {\bf 0506}, 029 (2005)
  [arXiv:hep-ph/0501072].
  %%CITATION = HEP-PH 0501072;%%
  
  %23
   %\cite{Dighe:1995bm}
\bibitem{Dighe:1995bm}
  A.~S.~Dighe,
  %``Determination of CKM phases through rigid polygons of flavor SU(3)
  %amplitudes,''
  Phys.\ Rev.\ D {\bf 54}, 2067 (1996)
  [arXiv:hep-ph/9509287];
  %%CITATION = HEP-PH 9509287;%%
  %\cite{Gronau:1995ng}
%\bibitem{Gronau:1995ng}
  M.~Gronau and J.~L.~Rosner,
  %``Determining the Weak Phase $\gamma$ From Charged $B$ Decays,''
  Phys.\ Rev.\ D {\bf 53}, 2516 (1996)
  [arXiv:hep-ph/9509325];
  %%CITATION = HEP-PH 9509325;%%
  %\cite{Grinstein:1996us}
%\bibitem{Grinstein:1996us}
  B.~Grinstein and R.~F.~Lebed,
  %``SU(3) Decomposition of Two-Body B Decay Amplitudes,''
  Phys.\ Rev.\ D {\bf 53}, 6344 (1996)
  [arXiv:hep-ph/9602218].
  %%CITATION = HEP-PH 9602218;%%
  
  %24
  %\cite{Chiang:2003rb}
\bibitem{Chiang:2003rb}
  C.~W.~Chiang, M.~Gronau and J.~L.~Rosner,
  %``Two-body charmless B decays involving eta and eta',''
  Phys.\ Rev.\ D {\bf 68}, 074012 (2003)
  [arXiv:hep-ph/0306021];
  %%CITATION = HEP-PH 0306021;%%
  %\cite{Chiang:2004nm}
%\bibitem{Chiang:2004nm}
  C.~W.~Chiang, M.~Gronau, J.~L.~Rosner and D.~A.~Suprun,
  %``Charmless B $\to$ P P decays using flavor SU(3) symmetry,''
  Phys.\ Rev.\ D {\bf 70}, 034020 (2004)
  [arXiv:hep-ph/0404073].
  %%CITATION = HEP-PH 0404073;%%
  
 %25
  %\cite{Williamson:2006hb}
\bibitem{Williamson:2006hb}
  A.~R.~Williamson and J.~Zupan,
  %``Two body B decays with isosinglet final states in SCET,''
  arXiv:hep-ph/0601214.
  %%CITATION = HEP-PH 0601214;%%
  %%Cited 0 times in SPIRES-HEP
 
%26
%\cite{Abe:2005iy}
\bibitem{Abe:2005iy}
  K.~Abe {\it et al.}  [Belle Collaboration],
  %``Improved measurements of branching fractions and CP asymmetries in B $\to$
  %eta h decays,''
  BELLE-CONF-0525, arXiv:hep-ex/0508030.
  %%CITATION = HEP-EX 0508030;%%
 
 %27
\bibitem{Savage:1989ub}
  M.~J.~Savage and M.~B.~Wise,
  %``SU(3) Predictions For Nonleptonic B Meson Decays,''
  Phys.\ Rev.\ D {\bf 39}, 3346 (1989)
  [Erratum-ibid.\ D {\bf 40}, 3127 (1989)].
  %%CITATION = PHRVA,D39,3346;%%

%28
\bibitem{Raz:2005hu}
  G.~Raz,
  %``Using SU(3) flavor to constrain the CP asymmetries in B $\to$ P P, V P, V V
  %decays involving b $\to$ s transitions,''
  arXiv:hep-ph/0509125.
  %%CITATION = HEP-PH 0509125;%%
  
\end{thebibliography}
\end{document}